# Structure of the normal state and origin of Schottky anomaly in the correlated heavy fermion superconductor UTe$_2$.

S. Khmelevskyi[1], L. V. Pourovskii[2,3] and E. A. Tereshina-Chitrova[4]

[1]*Research Center for Computational Materials Science and Engineering, Technical University of Vienna, Karlsplatz 13, A-1040, Vienna, Austria.*

[2]*Centre de Physique Théorique, Ecole Polytechnique, CNRS, Institut Polytechnique de Paris, 91128 Palaiseau Cedex, France*

[3]*Collège de France, 11 place Marcelin Berthelot, 75005 Paris, France*

[4]*Institute of Physics, Czech Academy of Sciences, 18121 Prague, Czech Republic*

**The newly discovered UTe$_2$ superconductor is regarded as a heavy fermion mixed-valence system with very peculiar properties within the normal and superconducting states. It shows no signs of magnetic order but strong anisotropy of a magnetic susceptibility and a superconducting critical field. In addition to the heavy fermion-like behavior in the normal state, it exhibits also a distinctive Schottky-type anomaly at about 12 K and a characteristic excitations gap ~35-40 K. Here we show, by virtue of dynamical mean-field theory calculations with a quasi-atomic treatment of electron correlations, that *ab-initio* derived crystal-field splitting of the 5f$^2$ ionic configuration yields an agreement with these experimental observations. We analyze the symmetry of magnetic and multipolar moment fluctuations that might lead to the superconducting pairing at low temperatures. A close analogy of the normal paramagnetic state of UTe$_2$ to that of URu$_2$Si$_2$ in the Kondo arrest scenario is revealed.**

Heavy fermion materials represent a natural platform for unconventional superconductivity, where the electron-electron correlations, magnetism and electron localization effects participate in a complex interplay in metals[1]. The heavy fermion compound $UTe_2$, in which the unconventional superconductivity has been recently discovered[2], attracts a considerable attention over the last few years[3] owning to the unusual properties of its superconducting and normal states. The superconducting (SC) state in $UTe_2$ is claimed to be a spin triplet[2] and is characterized by large and strongly anisotropic upper critical field and re-entrant superconducting dome at high applied magnetic field (~40 T) in a certain crystallographic direction[4]. Observations of chiral in-gap surface states[5] and broken time-reversal symmetry[6] suggest a highly non-trivial structure of the SC order parameters (OP). The pairing mechanism leading to the SC in $UTe_2$ still actively debated[7,8,9]. A particular problem for identification of the leading pairing mechanism is an intricate and a very rich heavy-fermionic and Kondo-lattice physics of the normal state[3] in $UTe_2$. The low temperature behavior of the resistivity and large value of specific heat point to a heavy fermion behavior[2,10]. The resistivity[10] and magnetic susceptibilities[11] are strongly anisotropic. Local maxima in the temperature dependence of resistivity suggest an emergence of Kondo hybridization[2]. There are no traces of any magnetic or "hidden order" phase transition down to the lowest temperatures. However, a broad Schottky-type anomaly in the temperature evolution of the specific heat with a maximum around 12-14 K[12,13] might be partially consistent with a characteristic energy scale observed in the inelastic neutron scattering (INS) experiments[14,15] and an characteristic energy gap[16]. It thus might imply the existence of two localized energy levels. Recently it was noted[9] that theoretical approaches for explaining the pairing symmetry in $UTe_2$ follow two general approaches. The first approach is to consider the Kramer's doublet at each of the U-sites forming narrow dispersive bands, while the symmetry of the SC OP is provided either by anisotropic ferromagnetic fluctuations[7] or by the band-anisotropy[8]. Another approach links the SC to the Kondo hybridization[9]. For a guided assessment of these proposals, it is essential to know the realistic structure of the UTe$_2$ normal state. Moreover, recent experiments[17] reveal the existence of two competing SC order parameters in UTe$_2$. Here we show that indeed two competing types of magnetic and orbital moments fluctuations with different symmetries are possible in UTe$_2$ as it follows from our *ab-initio* derived correlated electronic structure.

The electronic structure of UTe$_2$ has been investigated using different methodologies.[3] The standard local Density Functional Theory (DFT) calculations predict insulating ground state with a narrow gap and two narrow peaks below and above the Fermi energy with predominantly 5f-electrons character.[18] The metallic state is stabilized only by including effects of the single-site electron correlations in the 5f-shell, e.g by using straightforward DFT+U approximation[19,20] However, the spin-polarized DFT+U predicts a magnetically ordered ground state, inconsistent with experimental observations. More advanced methodology, like charge self-consistent exact diagonalization method, which models true paramagnetic state while allowing for on-site magnetic moment fluctuations, also predicts a metallic state.[21] The analysis of magnetic fluctuations derived with the later method has led the authors of Ref.[21] to the conclusion that 5f$^3$ configuration provides the leading contribution to the complex mixed-valence normal state of UTe$_2$. A mixed valence state with a major 5f$^3$-configuration weight was also suggested on the basis of a comparison of the UTe$_2$ experimental photoemission electron spectra with other U-based heavy electron intermetallic compounds[22]. However, recent experimental evidence contradicts the conclusions of Ref.[22] by showing that the 5f$^2$-configuration is dominating and its weight increases with the applied pressure.[17] Moreover, recent angle-resolved photoemission spectroscopy (ARPES) data were found to agree with ab-initio based DFT+dynamical mean-field theory (DMFT) calculations predicting a dominating (almost 84%) contribution of the 5f$^2$ $^3$H$_4$ multiplet into the U 5f ground state[23]. As we show below the structure of this multiplet is fully consistent with the experimental Schottky-like anomaly[12,13] and a characteristic energy scale found in INS experiments[14,15]. To derive the energy splitting and wave functions of the ionic U $5f^2$ configuration in the crystal field of UTe$_2$ we perform *ab initio* calculations employing a charge self-consistent DFT+DMFT[24,25] framework in conjunction with a quasi-atomic (Hubbard I) approximation[26] for correlation effects. This framework is abbreviated as DFT+HI below.

Our ab initio DFT+HI approach is based on a full-potential linear augmented plane waves (LAPW) band structure method[27] and a DMFT implementation provided by the "TRIQS" library[28,29]. The experimental UTe$_2$ lattice structure with the lattice parameter $a$=4.123 Å is used. The spin-orbit coupling is taken into account. 800 k-points are employed for the Brillouin-zone integration and the LAPW-basis cutoff parameter $R_{mt} \cdot K_{max}$ is set to 8. The Wannier orbitals for U 5$f$ shell are constructed using a projective approach of Ref. [30] from the Kohn-Sham eigenstates in the range [-1.02:2.72] eV around the Fermi level. Such a "narrow" projection window enclosing mainly U 5$f$ bands allows including the effect of hybridization on the crystal-field (CF) splitting within DFT+HI [35,31]. We verified that yet narrower window ([-1.02:2.04] eV) leads to a pathological DFT+HI electronic structure due to not well formed Wannier orbitals. We employ a rotationally invariant on-site Coulomb vertex specified by the parameters $F^0$=$U$=4.5 eV and $J_H$=0.6 eV, which are consistent with our previous studies of correlated uranium compounds[32]. The double-counting correction is calculated in the fully-localized limit[33] using the nominal occupancy of 5$f^2$ shell, as shown[34] to be appropriate for DFT+HI. The DFT+HI calculations were converged to 10$^{-5}$ Ry in the total energy. We employ the averaging scheme of Ref. [35] to suppress the self-interaction contribution to the crystal-field (CF) splitting, thus removing the CF splitting in the course of self-consistent DFT+HI iterations. Subsequently, the 5f levels and their energies within the obtained ground-state multiplet ($^3H_4$ of 5$f^2$) are evaluated from the converged DFT+HI one-electron level positions of the U 5$f$ shell.

The calculated *k*-resolved DFT+HI spectral function of the normal paramagnetic state of UTe$_2$ is shown in Fig, 1. As noted above, the predicted U 5*f* shell ground state is the $^3H_4$ ground-state multiplet with negligible mixing of the higher order multiplets of the 5f$^2$ configuration. The upper Hubbard band corresponding to an electron addition to this 5f$^2$ ground-state configuration forms a set of quasi-flat unoccupied bands with its bottom located about 0.25 eV above the Fermi energy (Fig. 1). The DFT+HI UTe$_2$ ground state is metallic, in agreement with the experiment. The metallicity provided by Tellurium p-bands crossing the Fermi level, which are strongly hybridized with the f-electron states above the Fermi level. The crystal environment of the U ion in UTe$_2$ splits the ground state $^3H_4$ multiplet into nine singlets with separation between the lowest and highest levels of about 970 K. We note that while the separation between 5*f*$^2$ ground state and the lowest energy 5*f*$^3$ is rather small (0.34 eV), it is still much larger than the total CF splitting of the $^3H_4$ manifold.

The calculated CF splitting of $^3H_4$ multiplet is displayed in Fig. 2. The two lowest energy levels has the energy splitting of 42 K that is in line with the characteristic energy scale of 35-40 K found[14,15] in various experiments, including the recent ARPES measurements[36]. In order to verify the consistency of the *ab-initio* derived energy splitting we calculate the temperature dependence of specific heat using all levels of the GS $^3H_4$ multiplet (from Fig. 2) and directly compare it to the available experimental estimations[12,13] in the Fig. 3. One can see a clear resulting Schottky anomaly with position of the maximum and overall behavior of the calculated *C/T* curve in excellent agreement with the electronic contribution derived from experiments. This implies that the physical scenario of dominating localized 5f$^2$ configuration in the electronic structure of the U-ion in the normal state of UTe$_2$ is a good starting point for the analysis of superconductivity and low-temperature properties in this compound.

To support further the last statement and to confirm quality of our CEF splitting calculations (Fig. 2), we compare our results with available experiments on low temperature magnetization and temperature dependent magnetic susceptibilities. The lowest energy states $|0>$ and $|1>$ allows for the states mixing only by the *a*-component of an applied external magnetic field component leading to the Ising type magnetic anisotropy along the *a*- crystallographic direction. Other components of magnetization might appear only due to mixing of higher energy levels in Fig. 2. This is in full agreement with a fact that the experimental magnetization easy axis is along the *a*-direction[2,11]. In the upper panel of Fig. 4 we show the calculated single-ion magnetization curves in different crystallographic directions at *T* = 1.8 K using the full multiplet structure of Fig. 2. As one can see the calculated magnetizations in overall good agreement with experiment also predicting the hard magnetic b-axis. The overestimation of the field induced magnetization observed along the intermediate c-axis might be due to some underestimation of the energy position of the third energy level (Fig. 2) due to inaccuracy of the Hubbard-I scheme that ignores small contribution of the $4f^3$ configuration to the ground state (< 16 %). However, comparison of the calculated temperature dependence of magnetization in 1 Tesla field and corresponding experimental curves[11] (lower panel of the Fig. 4) further demonstrate a good semi-quantitative agreement with experiment. In particular, the maximum observed in the temperature dependence of the susceptibility along the hard b-axis is well reproduced. Let's note, that our results are not applicable in the range of much higher applied fields, i.e. where the meta-magnetic transition is observed (~40 T)[3]. There a strong mixing of the $4f^3$ configuration can be expected, which according to our estimates is just about 0.3 eV above the $4f^2$ configuration. This mixing might be a source of the meta-magnetic transition in the hard magnetic direction.

To explore further consequences of our funding on possible understanding of superconductivity in UTe$_2$ we note the following analogy. A major contribution of the 4f$^2$ configuration has been also predicted[37] by full DFT+DMFT calculation in URu$_2$Si$_2$ superconductor, which features a famous "hidden-order" phase [38]. The DMFT calculations predict even higher weight of the 5f$^2$ configuration in the normal state of UTe$_2$ than in URu$_2$Si$_2$ (84 %[23] against ~65-70 %[37], respectively), hence suggesting a smaller tendency to the mixed valence in UTe$_2$ as compared to URu$_2$Si$_2$. However, both compounds share many common features in their normal state behavior. Due to the energy splitting of the lowest singlet states, similar in magnitude in both compounds, a Kondo physics[39] is developing at higher temperatures in both materials leading to similar temperature dependences of resistivity and magnetic susceptibility along different crystallographic axes[3,10,40]. At low temperatures the crystal field splitting of the ground state into separated singlet states leads to the Kondo arrest[37]. According to the Kondo arrest scenario[37] the inter-atomic exchange interactions mix the ionic localized singlets in a such way that the complex multipolar ordering, the so-called "hidden" order, occurs in UNi$_2$Si$_2$ at 17 K. Below this temperature strong magnetic fluctuations (or fluctuations of the dipole orbital moment operator) develop which mediate the superconducting pairing. The ultimate difference between UTe$_2$ and URu$_2$Si$_2$ is an absence of any traces of magnetic or "hidden" order phase transitions in UTe$_2$ down to the lowest temperatures. The situation, however, can be readily understood. In order to develop magnetic or "hidden" multipolar order in the system with the ground state quasi-doublet, the magnitude of the inter-atomic exchange interactions should exceed a certain critical value so that the two singlet states mix over the crystal field gap and the temperature effects do not destroy the order. On the mean-field level, the value of molecular field in the ordered state should be larger than exactly half of the crystal field gap[41]. This type of induced order in the systems with localized f-ions and with a singlet ground state is known as a "singlet" magnetism[42]. The reason for absence of phase transition in UTe$_2$ is that the total inter-atomic super-exchange coupling is not strong enough to overcome finite crystal field gap between singlets. In particular, it might be due to the structural peculiarity of $UTe_2$. The $UTe_2$ has a quasi 1-dimentional ladder type arrangement of U ions formed by connected U dimers with the dimer inter-atomic distance (3.78 Å) much shorter than the distance along the ladder edge (4.16 Å)[43]. Such a low coordination of U-ions in $UTe_2$ greatly reduces the overall strength of the two-site exchange and might bring additional low-dimensional effects by essentially increasing the interaction threshold value due to strong quantum fluctuations. Thus, the inter-atomic exchange interactions in $UTe_2$ are not necessarily smaller in magnitude than those in $URu_2Si_2$. They are just not strong enough to induce the magnetic dipole or a "hidden" multipolar order due to low nearest neighbor structural coordination and low structural dimensionality. However, they might be strong enough to induce critical fluctuations of orbital and magnetic dipole and multipole degrees of freedom at low temperatures and mediate the superconducting pairing. The evidence that $UTe_2$ is close vicinity to the critical threshold is also provided by the recent experiments that demonstrate[17] approaching to the quantum critical point at the applied pressure. In $URu_2Si_2$, the "hidden" order freezes fluctuations of even ranks of multipolar degrees of freedom, allowing only strongly anisotropic dipole magnetic fluctuations at low temperatures. In $UTe_2$, the critical magnetic dipole (rank 1) and multipolar fluctuations of both, odd and even, ranks are possible. It might also explain the competing SC order parameters recently reported in the experiments[17]. Before we begin to analyze the symmetry of possible fluctuations in $UTe_2$, let's note that the Kondo arrest scenario is fully consistent with the INS[14,15] and ARPES[36] observations of diffuse scattering in the 35-40 K range. At these temperatures, the Kondo effects emerge since both singlet states become almost equally populated allowing for the formation of the Kondo singlet with band states.

The calculated structure of the wave functions of the three lowest energy singlets (Fig. 2) in the |J,M> basis of $^3H_4$ multiplet is:

$$|0> = 0.281\,(|4;-4> + |4;+4>) + 0.636\,(|4;-2> + |4;+2>) - 0.177\,|4;0>,$$
$$|1> = 0.703\,(|4;-3> + |4;+3>) + 0.061\,(|4;-1> + |4;+1>), \qquad (1)$$

where the quantization axis for <J,M| (J = 4 is total orbital and M = {-4,..,4} – magnetic quantum numbers) basis is chosen to be along the c-crystallographic direction (direction of the shortest U-U bond) according a standard[3] setup for UTe$_2$. The two lowest singlet states can be mixed only by the inter-atomic exchange interactions between combinations of the orbital moment operator components that have non-zero matrix elements between these states. Despite that second exited singlet ( |2> ) in UTe$_2$ (see Fig.2) is just 20K above first one - |1> we restrict our analyzes to the lowest energy states since the interaction threshold where critical fluctuations would develops is ~10 K higher for the state |2> than for the state |1> and thus its importance for SC is less likely. In the following we use the tensorial operators based on real tesseral harmonics defined in the review by Santini *et al*[44]. Considering the lowest two states a non-zero matrix elements have: 1) odd rank dipolar $\widetilde{O}_1^1 = \hat{J}_x\,(\hat{J}_a)$, octupolar $\widetilde{O}_3^1$, and tricontadipolar $\widetilde{O}_5^5$, $\widetilde{O}_5^1$ moments; 2) even rank quadrupolar $\widetilde{O}_2^{-1}$ and hexadecapolar $\widetilde{O}_4^{-3}$ and $\widetilde{O}_4^{-1}$ moments. The even and odd rank fluctuations are independent of each other in a sense that they correspond to different singlet wave functions mixing schemes (with both real or complex coefficients, or one real and another one complex). But the fluctuations of the same parity will essentially induce fluctuations of the same parity. The unconventional superconductivity in the f-electron systems mediated by the high-rank multipolar moments is a known phenomenon, in particular, in systems with localized 5f configurations[45]. Hence, there are two types of strong fluctuations,

with potential to provide the SC pairing, allowed in UTe$_2$. These are the strongly anisotropic (odd parity) ones which break the time reversal symmetry and "nematic" (even parity) that do not break the time reversal symmetry. Being independent of each other, these fluctuations might compete leading to the two competing SC pairing schemes.

Thus, we show on the *ab-initio* level that domination of the 5f$^2$ configuration in low temperature normal state of UTe$_2$ revealed in earlier DFT+DMFT calculations[23] is fully consistent with experimentally observed specific heat anomaly, characteristic energy scale in the neutron diffraction and ARPES measurements[3]. In addition, it predicts well the anisotropy of magnetic response at low temperatures. The calculated structure of the lowest energy levels of U-ion explains the absence of magnetic or "hidden" order phase transitions down to the lowest temperatures and predicts the existence of two types of competing fluctuations with different symmetries which can provide the superconducting pairing.

Acknowledgements: We thank Dr. Y. Haga for helpful discussions on the procedure of the electronic specific heat subtraction in the experimental data, L. V. P. is grateful to the CPHT computer team for support. The work by E.A. T.-Ch. Is supported by the Czech Science Foundation (GAČR) under the grant number 22-19416S.

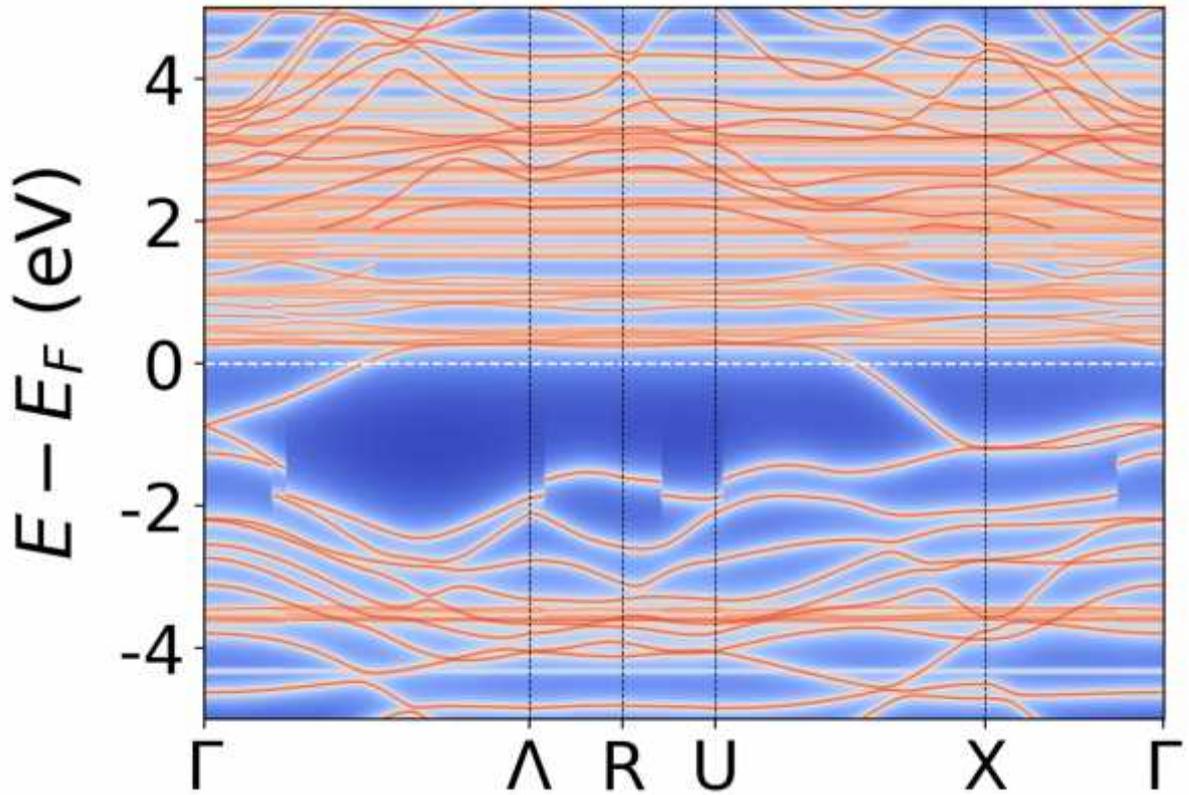

**Figure 1.** Calculated band structure (spectral density) of UTe$_2$ in the DFT+Hubbard I approximation with the localized 5f$^2$ configuration. A set of flat bands above the Fermi energy ($E_F$) is the 5f upper Hubbard band due to a transition from the $^3H_4$ multiplet of the U 5f$^2$ configuration to 5f$^3$ states. Metallic bands crossing the Fermi level are the p-bands of Te hybridized with the U 5f states. Small discontinuities seen in some band dispersions arise due to a different treatment of the KS states within and outside the projection window range, with the self-energy correction applied only to the KS bands within this window.

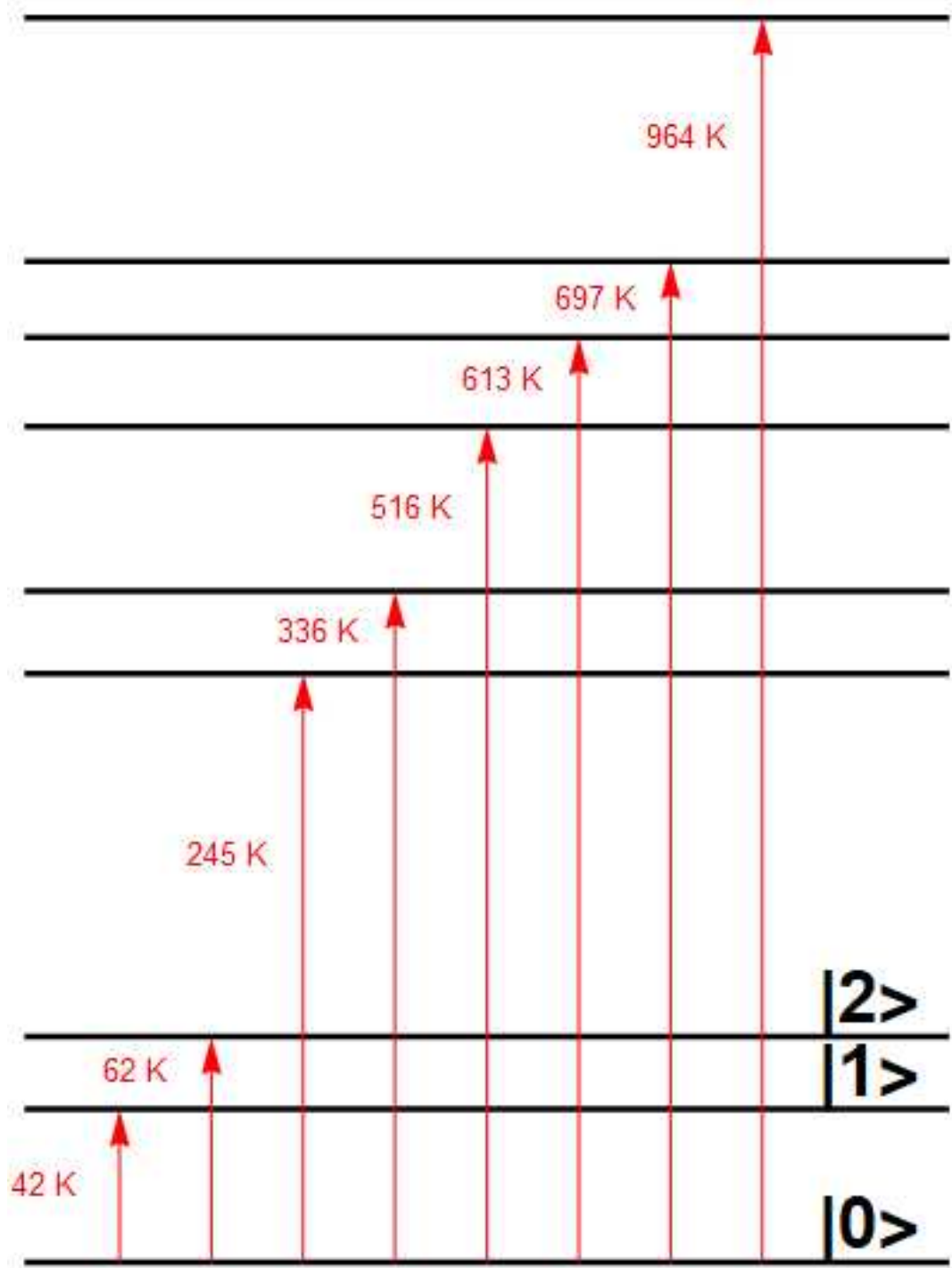

Figure 2. The calculated energy level splitting of the $^3H_4$ ground state multiplet of $5f^2$-configuration in UTe$_2$. The lower energy wave functions are given in the text in the |JM> basis representation.

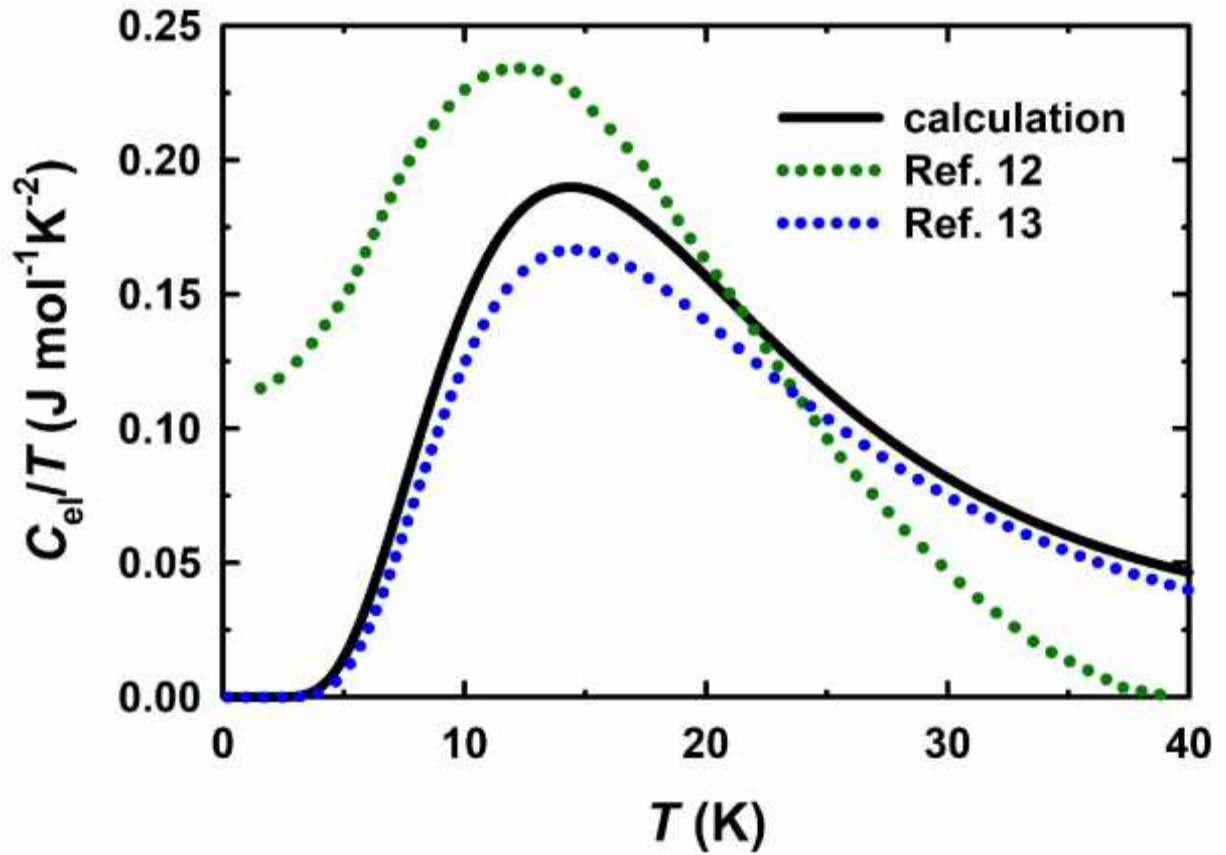

Figure 3. Comparison of temperature dependences of the calculated specific heat of $UTe_2$ and available experimental estimations. Black curve is a theoretical curve based on the *ab-initio* derived splitting from Fig. 2. Green and blue dots is the experimental data from Ref.[12] and Ref.[13], respectively. The differences between two experimental curves are due to different procedures used for subtraction the electronic part of the total specific heat (Y. Haga, private communication).

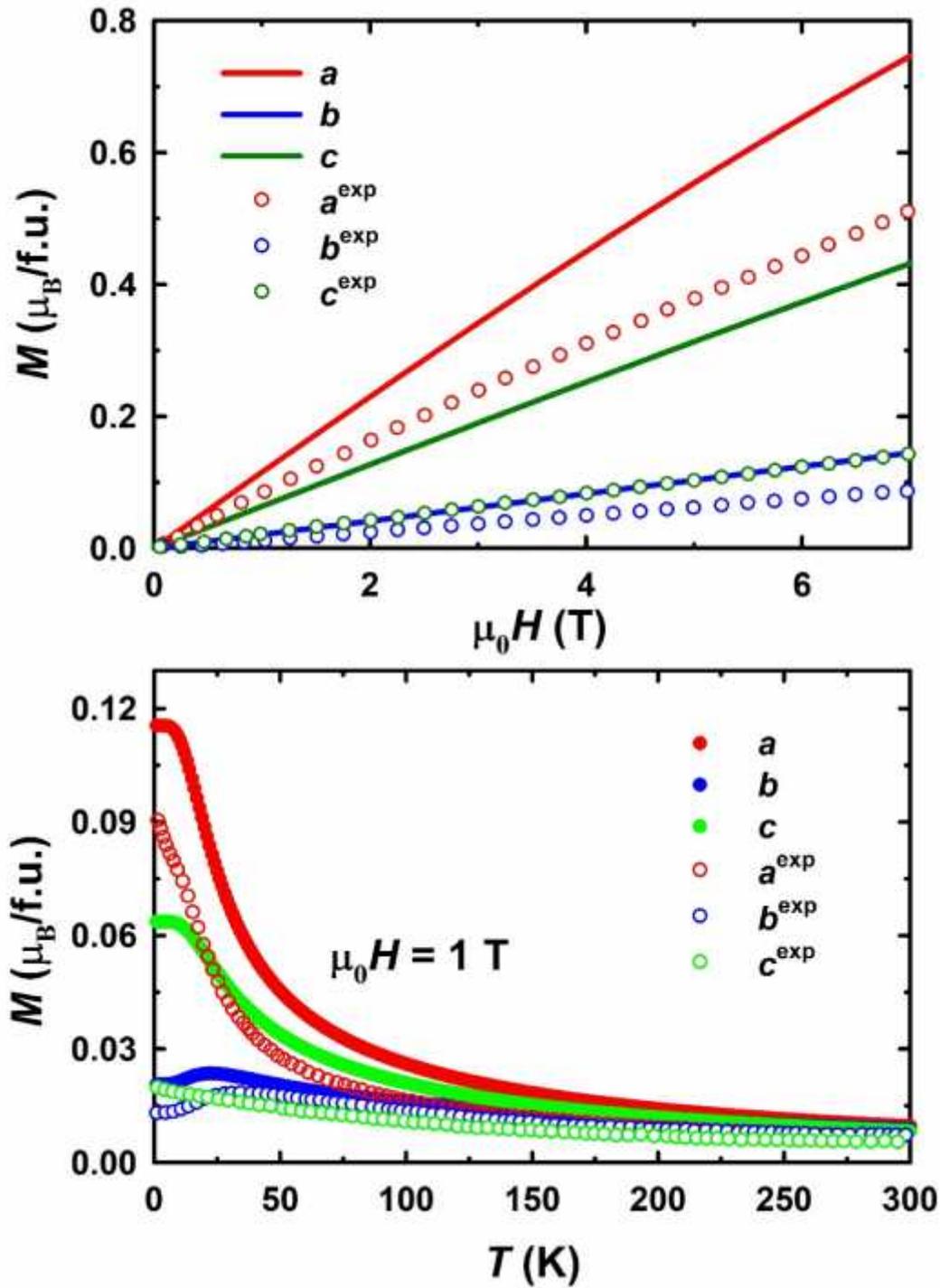

**Figure 4. (Upper panel)** Magnetization as a function of applied field at 1.8 K calculated (lines) for the three crystallographic axes and the experimental data (hollow symbols) adapted from Ran et al. [2]. **(Lower panel)** Comparison of the calculated and experimental (from Ref. 11) temperature dependences of magnetization for the field $\mu_0 H$ = 1 T applied along the three crystallographic axes.

[1] Q. Si and F. Steglich, Science **329**, 1161 (2010).

[2] S. Ran, C. Eckberg, Q.-P. Ding, Y. Furukawa, T. Metz, S. R. Saha, I.-L. Liu, M. Zic, H. Kim, J. Paglione, N. P. Butch, Science **365**, 684 (2019).

[3] D. Aoki, J.-P. Brison, J. Flouquet, K. Ishida, G. Knebel, Y. Tokunaga, Y. Yanase, J. Phys.: Condens. Matter **34**, 243002 (2022).

[4] G. Knebel, W. Knafo, A. Pourret, Q. Niu, M. Vališka, D. Braithwaite, G. Lapertot, M. Nardone, A. Zitouni, S. Mishra, I. Sheikin, G. Seyfarth, J.-P. Brison, D. Aoki, and J. Flouquet,, J. of Phys. Soc. Jap. 88, 063707 (2019).

[5] L. Jiao, S. Howard, S. Ran, Z. Wang, J. O. Rodriguez, M. Sigrist, Z. Wang, N. P. Butch, V. Madhavan, Nature **579**, 523 (2020).

[6] I. M. Hayes, D. S. Wei, T. Metz, J. Zhang, Y. S. Eo, S. Ran, S. R. Saha, J. Collini, N. P. Butch, D. F. Agterberg, A. Kapitulnik, J. Paglione, Science **373**, 797 (2021).

[7] J. Ishizuka and Y. Yanase, Phys. Rev. B **103**, 094504 (2021).

[8] T. Shishidou, H. G. Suh, P. M. R. Brydon, M. Weinert, and D. F. Agterberg, Phys. Rev. B **103**, 104504

[9] T. Hazra and P. Coleman, https://doi.org/10.48550/arXiv.2205.13529 (2022).

[10] Y.S. Eo, Sh. Liu, Sh. R. Saha, H. Kim, Sh. Ran, J. A. Horn, H. Hodovanets, J. Collini, T. Metz, W. T. Fuhrman, A.H. Nevidomskyy, J.D. Denlinger, N.P. Butch, M.S. Fuhrer, L. Andrew Wray, J. Paglione, Phys. Rev. B **106**, L060505 (2022).

[11] A. Miyake, Y. Shimizu, Y. J. Sato, D. Li, A. Nakamura, Y. Homma, F. Honda., J. Flouquet, M. Tokunaga, and D. Aoki, J. Phys. Soc. Jap. **88**, 063706 (2019)

[12] K. Willa, F. Hardy, D. Aoki, D. Li, P. Wiecki, G. Lapertot, and C. Meingast, Phys. Rev. B **104**, 205107 (2021).

[13] Y. Haga, P. Opletal, Y. Tokiwa, E. Yamamoto, Y. Tokunaga, S. Kambe, and H Sakai, J. Phys.: Condens. Matter **34**, 175601 (2022). Y. Haga, private communication, July 2022.

[14] W. Knafo, G. Knebel, P. Steffens, K. Kaneko, A. Rosuel, J.-P. Brison, J. Flouquet, D. Aoki, G. Lapertot, S. Raymond, Phys. Rev. B **104**, L100409 (2021).

[15] N. P. Butch, S. Ran, S. R. Saha, P. M. Neves, M.P. Zic, J. Paglione, S. Gladchenko, Q. Ye, J. A. Rodriguez-Rivera, npj Quantum Mater. **7**, 39 (2022).

[16] Y. Tokunaga, H. Sakai, S. Kambe, Y. Haga, Y. Tokiwa, P. Opletal, H. Fujibayashi, K. Kinjo, S. Kitagawa, K. Ishida, A. Nakamura, Y. Shimizu, Y. Homma, D. Li, F. Honda, and D. Aoki, J. Phys. Soc. Jpn. **91**, 023707 (2022).

[17] S. M. Thomas, F.B. Santos, M. H. Christensen, T. Asaba, F. Ronning, J.D. Thompson, E. D. Bauer, R. M. Fernandes, G. Fabbris, and P. F. S. Rosa, Sci. Adv. **6**, eabc8709 (2020).

[18] D. Aoki, A. Nakamura, F. Honda, D. Li, Y. Homma, Y. Shimizu, Y. J. Sato, G. Knebel, J. P. Brison, and A. Pourret, D. Braithwaite, G. Lapertot, Q. Niu, M. Vališka, H. Harima, and J. Flouquet, J. Phys. Soc. Jpn. **88**, 043702 (2019).

[19] J. Ishizuka, S. Sumita, A. Daido, and Y. Yanase. Phys. Rev. Lett. **123,** 217001 (2019).

[20] A. B. Shick, and W. E. Pickett, Phys. Rev. B **100**, 134502 (2019).

[21] A. B. Shick, S. Fujimori, and W. E. Pickett, Phys. Rev. B **103**, 125136 (2021).

[22] S.-I. Fujimori, I. Kawasaki, Y. Takeda, H. Yamagami, A. Nakamura, Y. Homma, and D. Aoki, J. Phys. Soc. Jpn. **90**, 015002 (2021).

[23] L. Miao, S. Liu, Y. Xu, E. C. Kotta, C.-J. Kang, S. Ran, J. Paglione, G. Kotliar, N. P. Butch, J. D. Denlinger, and L. A. Wray, Phys. Rev. Lett. **124**, 076401 (2020).

[24] A. Georges, G. Kotliar, W. Krauth, and M. J. Rozenberg, Rev. Mod. Phys. **68**, 13 (1996).

[25] V. I. Anisimov, A. I. Poteryaev, M. A. Korotin, A. O. Anokhin, and G. Kotliar, J. Phys. Condens. Matter **9**, 7359 (1997).

[26] J. Hubbard, Proc. R. Soc. A **276**, 238 (1963).

[27] P. Blaha, K. Schwarz, G. Madsen, D. Kvasnicka, J. Luitz, R.Laskowski, F. Tran, and L. D. Marks, WIEN2k, An Augmented Plane Wave + Local Orbitals Program for Calculating Crystal Properties (Karlheinz Schwarz, Technische Universität Wien, Austria, 2018),

[28] O. Parcollet, M. Ferrero, T. Ayral, H. Hafermann, I. Krivenko, L. Messio, and P. Seth, Comput. Phys. Commun. 196, **398** (2015).

[29] M. Aichhorn, L. Pourovskii, P. Seth, V. Vildosola, M. Zingl, O. E. Peil, X. Deng, J. Mravlje, G. J. Kraberger, C. Martins *et al.,* Comput. Phys. Commun. **204**, 200 (2016).

[30] M. Aichhorn, L. Pourovskii, V. Vildosola, M. Ferrero, O.Parcollet, T. Miyake, A. Georges, and S. Biermann, Phys. Rev. B **80**, 085101 (2009).

[31] L. V. Pourovskii J. Boust, R. Ballou, G. Gomez Eslava, and D. Givord. Phys. Rev. B **101**, 214433 (2020)

[32] L. V. Pourovskii, and S. Khmelevskyi, Phys. Rev. B **99**, 094439 (2019).

[33] M. T. Czyżyk and G. A. Sawatzky, Phys. Rev. B **49**, 14211 (1994).

[34] L. V. Pourovskii, B. Amadon, S. Biermann, and A. Georges, Phys. Rev. B **76**, 235101 (2007).

[35] P. Delange, S. Biermann, T. Miyake, and L. Pourovskii, Phys. Rev. B **96**, 155132 (2017).

[36] S. M. Mekonen, C. Jong Kang, D. Chaudhuri, D. Barbalas, S. Ran, G. Kotliar, N. P. Butch, and N. P. Armitage, arXiv:2105.05121v2 (2022).

[37] K. Haule and G. Kotliar, Nature Phys. **5**, 796 (2009).

[38] J. A. Mydosh, and P. P. M. Oppeneer, Rev. Mod. Phys. **83**, 1301 (2011).

[39] D. L. Cox, Phys. Rev. Lett. **59**, 1240 (1987).

[40] T. T. M. Palstra, A. A. Menovsky, J. van den Berg, A. J. Dirkmaat, P. H. Kes, G. J. Nieuwenhuys, and J. A. Mydosh, Phys. Rev. Lett. **55,** 2727 (1985).

[41] K. Haule and G. Kotliar, Eur. Phys. Lett. **89**, 57006 (2010).

[42] B. Bleaney, Proc. Roy. Soc. A279, **19** (1963).

[43] K. Stöwe, J. Alloys and Compounds **246**, 111 (1997).

[44] P. Santini, S. Carretta, G. Amoretti, R. Caciuffo, N. Magnani, and G. H. Lander, Rev. Mod. Phys. **81**, 807 (2009).

[45] M. Tsujimoto, Y. Matsumoto, T. Tomita, A. Sakai, and S. Nakatsuji, Phys. Rev. Lett. **113**, 267001 (2014).